\documentclass{siamltex}
\usepackage{graphicx}

\title{Delayed-mutual coupling dynamics of lasers: scaling laws and resonances}
\author{T.W. Carr
\thanks{
Department of Mathematics,
Southern Methodist University,
Dallas, TX 75275-0156,
tel: 214-768-3460,
tcarr@smu.edu}
\and
I.B. Schwartz
\thanks{
Nonlinear Dynamical Systems Section, Code 6792,
Naval Research Laboratory,
Washington, DC 20375,
tel: 202-404-8359,schwartz@nlschaos.nrl.navy.mil}
\and
Min-Young Kim
\thanks{
Department of Physics,
University of Maryland,
College Park, MD, 20742,
tel: 514-398-8224,
mmyykim@cnd.mcgill.ca
}
\and
Rajarshi Roy
\thanks{
Department of Physics,
University of Maryland,
College Park, MD, 20742,
rroy@glue.umd.edu
}
}

\begin{document}

\maketitle
\begin{abstract}We consider a model for two lasers that are mutually coupled optoelectronically
by modulating the pump of one laser with the intensity deviations
of the other. Signal propagation time in the optoelectronic loop
causes a significant delay leading to the onset of oscillatory output. Multiscale
perturbation methods are used to describe the amplitude and period of oscillations
as a function of the coupling strength and delay time. For weak coupling the oscillations
have the laser's relaxation period, and the amplitude varies as the one-fourth power
of the parameter deviations from the bifurcation point. For
order-one coupling strength the period is determined as multiples of the delay time,
and the amplitude varies with a square-root power law.
Because we allow for independent
control of the individual coupling constants, for certain parameter values there is
an atypical amplitude-resonance phenomena. Finally, our theoretical results are consistent
with recent experimental observations when the inclusion of a low-pass filter in the
coupling loop is taken into account.
\end{abstract}

%*******************************************************************************
%*******************************************************************************
%*******************************************************************************
\section{Introduction}

In recent work, we presented experimental and simulation results
for two mutually coupled lasers \cite{KiRoArCaSc05} with time-delayed asymmetric
coupling. The light emitted from one laser propagates through fiber-optic
cables to a photodetector that generates an electronic signal proportional
to the light-intensity deviations from steady state. The electronic signal
may be attenuated or amplified before it modulates the pump current of the other
laser. The propagation time of the signal in the optoelectronic path
introduces a significant time delay, and the coupling strength in each direction
can be controlled separately. Our work in \cite{KiRoArCaSc05}
investigated how the time delay and asymmetric coupling led to oscillatory
and pulsating laser output. In this paper, we present a
more theoretical exploration of the dynamics that result from this
coupling configuration.

In \cite{CaTaSc05} we investigated this same configuration
(asymmetric pump coupling) but without including the effect of delay.
In addition, the coupling constant from laser-2 to laser-1 was negative,
while the coupling constant from laser-1 to laser-2 was positive.
That is, we assumed that the electronic coupling signal from laser-2 was
inverted before applying it to the pump of laser-1. For purely
harmonic signals, having opposite-sign coupling constants is equivalent to time-delayed
coupling when the delay is half of the period. We found that
as the coupling constant from laser-2 to laser-1 is increased in magnitude:
(i) There is a Hopf bifurcation to oscillatory output. (ii) For certain
parameter values there exists a small-coupling resonance such that the amplitudes
of both lasers are pulsating. (iii) For large coupling, laser-2 acts as
a small-amplitude, nearly harmonic modulation to laser-1. Laser-1 exhibits
period-doubling bifurcations to chaos and complex subharmonic resonances.

In this paper we include the effect of the delayed-coupling, which results
in a coupled set of delay-differential equations (DDEs). With delay there is again
a Hopf bifurcation to oscillatory output. However, the delay allows for periodic
oscillations not just at the laser's relaxation period as in \cite{CaTaSc05},
but at periods that are integer multiples of the delay; we refer to the
former oscillatory solutions as ``internal modes," and the latter as
``external modes." We show that as the coupling strength is increased,
the first instability is a Hopf bifurcation to the internal mode. Our nonlinear
analysis then shows that the amplitude of the internal mode varies as the
one-fourth root of the bifurcation parameter's deviation from the Hopf bifurcation point.

However, in the experiments of \cite{KiRoArCaSc05} the output was not
the internal mode but an external mode. The period of the oscillatory output
was a multiple of the delay, and the amplitude of the
oscillations varied as the square root of the bifurcation parameter.
The discrepancy is due to the fact that the optoelectronic coupling contains
an intrinsic low-pass filter that attenuates the
internal mode. We show that the filter selects the external mode with the greatest
possible frequency passed by the low-pass filter. A nonlinear analysis
then shows that the amplitude of the external modes does indeed vary
with a square-root power law.

For the coupled lasers without delay in \cite{KiRoArCaSc05}, we found that
the parameters could be tuned to cause a resonance-type effect with respect
to the coupling; more specifically, the coupling could be tuned to
maximize the amplitude of the oscillatory output. We find that this resonance effect
also occurs with the inclusion of the delay. However, there are some
dramatic qualitative differences as the coupling parameters are varied.
In the system with delay, the branch of periodic solutions in the bifurcation
diagram can be smoothly folded to form parameter intervals of bi-stability. Then
there is a critical value of the same coupling parameter beyond
which the bi-stability disappears suddenly and non-smoothly.

Our coupling scheme is an example of ``incoherent coupling''
\cite{YaKa88} because it depends only on the laser's intensity and not on
the complex electric field. This is because the intensity of one laser
affects the other only indirectly; in our case, the intensity of one laser
modulates the pump current of the other.
In contrast, ``coherent coupling'' refers to when the optical field
of one laser is directly injected into the cavity of the other laser.
Analogously, a single laser with delayed and re-injected self-feedback
would be called ``coherent'' feedback. Semiconductor lasers with
delayed coherent feedback have been extensively studied because of
the their widespread application in electronics and communication systems;
there the laser's output signal may be reflected off external surfaces
back into the laser. The Lang-Kobayashi DDEs \cite{LaKo80} are the
canonical model for semiconductor lasers with delayed self-feedback and have been
used to investigate phenomena ranging from the onset of instabilities
to the development of chaotic output referred to as ``coherence collapse''
(see \cite{VaLe95} for a review). Coupled sets of Lang-Kobayashi or
related DDEs are most often used to described coherently-coupled semiconductor
lasers with delay. Two recent studies \cite{JaMaPi03,ErKrLe06} are in
the same spirit as this paper as they track the number of and
properties of oscillatory solutions that appear as the coupling strength
is increased.

The main simplification that results from considering incoherent coupling
between the lasers is that neither the laser-frequency detuning nor the electric-field
phase differences affect the dynamics of the coupled system \cite{LiMaEr90}.
Pieroux et al. \cite{PiErOt94} have shown that the electrooptical coupling we
consider in this paper leads to an equivalent dynamical model as incoherent coupling.
There have recently
been a number of other investigations of lasers with incoherent coupling.
The main focus of many of these works was on chaotic synchronization
(see \cite{SuGaEr05} and included references).Two recent papers by
Vicente et al. \cite{ViTaMu04,ViTaMu06} consider a similar implementation
of optoelectronic coupling to that we consider here, i.e., the intensity of
one laser modulates the pump current of the other. Their investigation of
the oscillatory solutions is mainly numerical and they discuss interesting
phenomena such as amplitude quenching, frequency locking and routes to chaos.

Delay-coupled relaxation and limit-cycle oscillators have been the subject
of many investigations (see \cite{ReSeJo03,SeRa03,WiRa02,ReSeJo98,CaWa98,LeKiKo97} and
their included references). These systems are self-oscillatory and
the amplitude is often fixed by the intrinsic properties
of the oscillator. In many cases the system may be reduced
to a phase equation, or a system of phase equations if the oscillators
are coupled \cite{YeSt99}.

In contrast, for laser systems the amplitude is an important dynamical
variable. This is because the lasers we consider are weakly damped, nearly conservative
systems such that the steady state is the only asymptotically stable
solution. Oscillations must be induced by external mechanisms such
as modulation, injection or coupling \cite{AbMaNa88}. Thus, the amplitude
of the oscillations is highly dependent upon the external mechanism,
in our case the coupling, rather than the individual laser.
We should mention that for coupled limit-cycle oscillators there is
the amplitude instability referred to as ``amplitude death,''
where for specific values of the coupling (and oscillator parameter values)
the amplitude becomes zero.

After presenting our model, we show that the linear-stability analysis
predicts a critical value of the coupling constant (that depends upon
the delay time) such that there is Hopf bifurcation to the internal mode,
i.e., oscillations with the intrinsic relaxation period.
We then use multiple-scale perturbation methods, modified to account
for time delays \cite{PiErGaKo00}, to analyze the long-time evolution of
the internal mode. The results are a pair of complex Stuart-Landau
DDEs for the oscillation
amplitudes that include a slow-time delay term. We analyze the
amplitude equations to determine the bifurcation properties of the internal mode.
Of particular note is that we allow for independent control of the
coupling constants; most other studies of coupled lasers and
oscillators consider symmetric coupling, where the coupling is the same
across all of the elements. The independent control is important
because, as we will
show, it allows for a singularity in the bifurcation equations
that marks the ``resonance" of large amplitude solutions.

The Hopf bifurcation to the internal mode occurs for small coupling. However,
as the coupling is increased, the external modes appear via Hopf bifurcations.
We have extended our multiple-scale analysis to be able to describe the bifurcation of the
external modes that occur for $O(1)$ coupling. The analytical challenge is that the
delay terms remain a part of the leading-order problem. Describing
the bifurcation of the external modes is important for
comparing our results to those in experiments \cite{Ki05},
because the experimental system contains low-pass filters that attenuate the
internal mode originating from the first Hopf bifurcation.

An alternative to our multiple-scale method for deriving slow-time
amplitude equations is to apply center-manifold theory with averaging
\cite{ChFe04,SeRa03,WiRa02}. However, the averaging method does not
retain the delay in the slow time; that is, the amplitude equations
are ordinary-differential equations, not DDEs. The distinction is
not important for their or our investigation because we look for
steady-state solutions of the amplitude equations. However, time-varying
amplitudes may require consideration of the slow-time delay.

Reddy et al. \cite{ReSeJo98,ReSeJo99} considered delay-coupled
Stuart-Landau DDEs similar to the amplitude equations we derive here.
However, the complex coefficients they use are appropriate for
limit-cycle oscillators and not the lasers that we consider.
They focus their work on the properties of synchronization and
amplitude death.

In the next section, we nondimensionalize the model for two-coupled
lasers, with the result being the focus of the rest of the paper. We perform the
linear-stability analysis in Sec.~\ref{s:linstab}. Slow-time evolution equations
for small and $O(1)$ coupling are derived in Sec.~\ref{s:intmodes} and
Sec.~\ref{s:extmodes}, respectively. We close with a discussion of the
results.

%*******************************************************************************
%*******************************************************************************
%*******************************************************************************

\section{Class-B model}

We consider two class-B lasers \cite{AbMaNa88,ArLiPuTr84} modeled by
rate equations as
 \begin{eqnarray}
\frac{dI_{j}}{dt} & = & (D_{j}-1)I_{j},\quad j=1,2 \nonumber\\
\frac{dD_{j}}{dt} & = & \epsilon_{j}^2[A_{j}(t)-(1+I_{j})D_{j}],
\label{eq:IDmodel}
\end{eqnarray}
where $I_j$ and $D_j$ are the scaled intensity and population inversion, respectively,
of each laser. Scaled time $t$ is measured with respect to the cavity-decay constant $\kappa$, $t=(2\kappa)\: t_r$, where $t_r$
measures real time. $\epsilon^2 = \gamma_{\parallel}/(2\kappa)$ is a ratio of the
inversion-decay rate, $\gamma_{\parallel}$, to the cavity-decay rate, $\kappa$,
and measures the relative life-time of photons to excited electrons. $A(t)$ is the
dimensionless pump rate and corresponds to the energy input to the
laser by an external source, e.g., another laser, an incoherent light
source or an electronic current. The mass-action coupling term $ID$ models
``stimulated emission"; a photon passing through the laser cavity
stimulates an excited electron to drop to the lower energy level, resulting
in one less excited electron and one more photon. Eqs.~(\ref{eq:IDmodel})
may be derived as a reduced model from semi-classical Maxwell-Bloch equations
\cite{AbMaNa88,ArLiPuTr84,Ma97}; the latter use Maxwell's equations to describe the
laser's electromagnetic fields coupled with the Bloch equations from quantum
mechanics for the amplifying media.

If the pump rate is a constant, $A_j(t) = A_{j0}$, then the laser relaxes
to the steady state $D_{j0}=1$, $I_{j0}=A_{j0}-1$.
To facilitate further analysis, we define new variables for the deviations
from the non-zero steady state \cite{ScEr94} as
\begin{equation}
	I_{j}=I_{j0}(1+y_{j}),\quad D_{j}=1+\epsilon_{j}\sqrt{I_{j0}}x_{j},\quad
	s=\epsilon_{1}\sqrt{I_{10}}t.
\end{equation}
Our goal is to investigate the effects of coupling through the pump.
In addition, we account for the effect of a delay when the signal from
one laser takes a non-negligible time before affecting the other. Thus, the
pump coupling is taken to be
\begin{equation}
	A_{j}(t)=A_{j0}-I_{j0}\delta_{k}y_{k}( t - \tau_k ) .
	\label{eq:negcoupling}
\end{equation}
Thus, we feed the intensity \textit{deviations} $y_k = (I_k-I_{k0})/I_{k0}$ from the
steady-state output of laser $k$ to the pump of laser $j$; the strength of the coupling is
controlled by $\delta_{k}$. The pump-coupling scheme allows for easy electronic
control of the coupling signal. The negative coupling results from the configuration
of the electronic coupling circuits in \cite{KiRoArCaSc05}. Finally, we assume
that the decay constants of the two lasers are related by
$\epsilon_2=\epsilon_{1}\frac{\sqrt{I_{10}}}{\sqrt{I_{20}}}\beta.$
The new rescaled equations are
\begin{eqnarray}
	\frac{dy_{1}}{dt} & = & x_{1}(1+y_{1}),\label{eq:xymodel}\nonumber\\
	\frac{dx_{1}}{dt} & = & -y_{1}-\epsilon x_{1}(a_{1}+by_{1})-\delta_2y_2(t-\tau_2),\nonumber\\
	\frac{dy_2}{dt} & = & \beta x_2(1+y_2),\nonumber\\
	\frac{dx_2}{dt} & = & \beta[-y_2-\epsilon\beta x_2(a_2+by_2)-\delta_{1}y_{1}(t-\tau_1)],
	\label{e:laserxy}
\end{eqnarray}
where
\begin{equation}
	a_{1}=\frac{1+I_{10}}{\sqrt{I_{10}}},\quad
	a_2=\frac{\sqrt{I_{10}}(1+I_{20})}{I_{20}},\quad \mbox{and }
	b=\sqrt{I_{10}}.
\end{equation}
For notational convenience we have let $s\rightarrow t$ and dropped the subscript on
$\epsilon_{1}$($\epsilon_{1}\rightarrow\epsilon)$. Rogister et al. \cite{RoPiSc02} have
considered a very similar model to ours; however, in their case the cross-coupling
term is instantaneous, while the delay appears through self-feedback of each laser's
own intensity.

%*******************************************************************************
%*******************************************************************************
%*******************************************************************************

\section{Linear stability of the steady state}
\label{s:linstab}

%*******************************************************************************
%*******************************************************************************
\subsection{Characteristic equation}

In the new variables, the steady state is given by $x_j = y_j = 0$. The linear
stability of the steady state is determined by studying the evolution of small
perturbations, for which we obtain the characteristic equation
\begin{equation}
  [ \lambda ( \lambda + \epsilon a_1 ) + 1 ] [ \lambda ( \lambda + \epsilon
  a_2 \beta^2 ) + \beta^2 ] - \beta^2 \delta_1 \delta_2 e^{- \lambda ( \tau_1
  + \tau_2 )} = 0.
\end{equation}
The delay results in the exponential term $\exp ( - \lambda ( \tau_1 + \tau_2)
)$. The transcendental form of the characteristic equation and, hence, the
possibility of an infinite number of roots is typical for DDEs.

If either $\delta_j = 0$, then there are only decaying oscillations, which
indicates that a single uncoupled laser is a weakly damped oscillator; this is the
general characteristic of ``class-B'' lasers \cite{ArLiPuTr84}. For $\delta_j \ne 0$ numerical
simulations indicate there is a Hopf bifurcation as the $\delta_j$ are
increased. To identify the Hopf bifurcation point, we let $\lambda = i \omega$
and separate the characteristic equation into real and imaginary parts. After
some algebra we obtain a single equation for the frequency and another
equation for the value of $\Delta = \delta_1 \delta_2$ at the bifurcation
point. The equations are
\begin{eqnarray}
  0 & = & \tan ( \omega \tau_s ) [ ( 1 - \omega^2 ) ( \beta^2 - \omega^2 ) -
  \epsilon^2 a_1 a_2 \beta^2 \omega^2 ] \\
    &&  + \epsilon \omega ( 1 - \omega^2 ) (
  a_1 + a_2 \beta^2 ), \nonumber \\
  \beta^2 \Delta_H^2 & = & [ ( 1 - \omega^2 ) ( \beta^2 - \omega^2 ) \label{e:DH}\\
   && - \epsilon^2 a_1 a_2 \beta^2 \omega^2 ]^2 + [ \epsilon \omega ( 1 - \omega^2 )
  ( a_1 + a_2 \beta^2 ) ]^2 \nonumber ,
\end{eqnarray}
where $\tau_s = \tau_1 + \tau_2$ is the ``round-trip'' delay time. The
transcendental equation for $\omega$ can be solved numerically and its
solution is then substituted into Eq.~(\ref{e:DH}) to determine the value of the
coupling at the bifurcation point $\Delta = \Delta_H$.

To simplify further discussion we consider $\beta =1$. This implies
$\epsilon_2 = \epsilon_1 \sqrt{I_{10}/I_{20}} $, which is a specific
relationship between the lasers' decay constants, $\gamma_{\parallel,j}$ and $\kappa_j$,
and the pump rate, $A_j0$; if $\beta = 1$ and $A_{10} = A_{20}$ then the lasers are
identical. The results are qualitatively the same for nearly identical lasers
with $\beta \approx 1$.
To simplify notation, we define $c_1 = a_1 + a_2$ and $c_2 = a_1 a_2$ and have
\begin{eqnarray}
  0 & = & \tan ( \omega \tau_s ) [ ( 1 - \omega^2 )^2 - \epsilon^2 c_2
  \omega^2 ]
     + \epsilon \omega ( 1 - \omega^2 ) c_1,  \label{eq:omega}\\
  \Delta_H^2 & = & [ ( 1 - \omega^2 )^2 - \epsilon^2 c_2 \omega^2 ]^2
      + [\epsilon \omega ( 1 - \omega^2 ) c_1 ]^2,  \label{eq:DH}
\end{eqnarray}
For $\epsilon \ll 1$ the leading approximation to the frequency is given by
$\omega = \pm 1$ or $\omega = m \pi / \tau_s$, $m$ an integer. We will refer
to the former as the \textit{internal mode} because this is the scaled laser-relaxation
frequency; more specifically, an oscillatory solution of Eq.~(\ref{e:laserxy}) with
period $\omega = 1$ is called the internal mode. Similarly, oscillatory
solutions with period $\omega = m \pi / \tau_s$, $m$ an integer, are
called \textit{external modes} because their periods are determined by the delay.
For any fixed value of the delay $\tau_s$ there is the internal mode and an
infinite number of external modes.

\begin{figure}
\begin{center}
\includegraphics[width=4in]{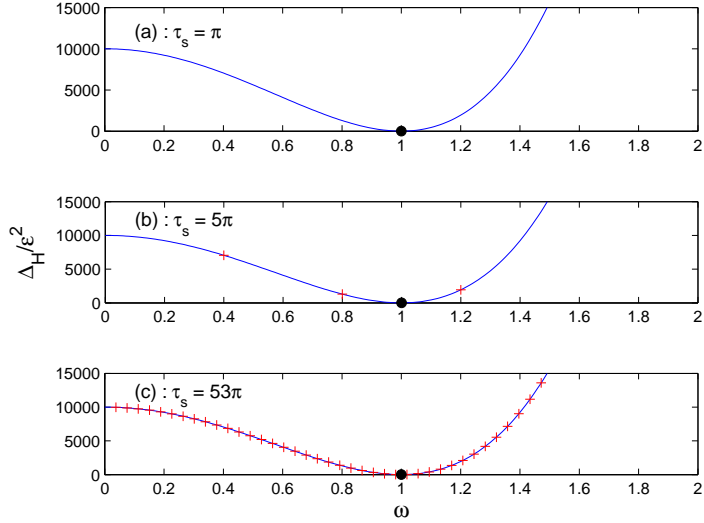}
\end{center}
\caption{Neutral stability curve $\Delta_H(\omega)$ (solid curve) for Hopf
bifurcations described by Eq.~(\ref{eq:DH}). The solid circle at
$\omega \approx 1$ indicates the internal mode. The crosses (+) at
$\omega \approx m\pi/\tau_s$ indicate the external modes. Because
$\tau_s = n\pi$, then $\omega \approx m/n$ for the external modes.
($\epsilon = 0.01, a_1 = a_2 = 2, b=1, \beta =1$)}
\label{f:deltavsomega}
\end{figure}

From Eq.~(\ref{eq:DH}) we see that for almost all values of the delay,
as the coupling is increased from zero, the bifurcation to the internal mode
occurs at a lower value of the $\Delta_H$ than
the external mode (this is true except when $\tau_s = n\pi$, $n$ even,
which we will discuss in later paragraphs). More specifically, for the
internal mode with $\omega \approx 1$ the Hopf bifurcation occurs when
$\Delta_H = O ( \epsilon^2 ) \ll 1$. For the external
mode with $\omega = O ( 1 )$ the value of the coupling at the Hopf
bifurcation is $\Delta_H = O ( 1 )$. This is illustrated in
Fig.~\ref{f:deltavsomega}, where we
have plotted Eq.~(\ref{eq:DH}). We will determine more precise approximations
for these bifurcation points below; however, we will first make additional
qualitative observations.

Let us fix the delay to be a multiple of $\pi$, i.e., $\tau_s = n\pi$, $n$
an integer. Then $\omega = \pm 1$, corresponding to the internal mode,
is an exact solution to the
frequency equation Eq.~(\ref{eq:omega}) and $\Delta_H = \pm \epsilon^2 c_2$.
The $+$ is taken if $n$ is odd and the $-$ is taken if $n$ is even
(the sign association comes from the real and imaginary parts of the
characteristic equation before $\omega$ and $\Delta$ were separated).
In the present paper, we will consider $\delta_j > 0$ and $\Delta > 0$;
results for $\Delta < 0$ are qualitatively the same. Thus, we have the exact solution
$\omega = \pm 1$, $\Delta_H = + \epsilon^2 c_2$ when $\tau_s = n \pi$,
$n$ odd. An analysis of Eq.~(\ref{eq:DH}) shows this occurs at a minimum
in the curve $\Delta_H(\omega)$ as illustrated in Fig.~\ref{f:deltavsomega}.

If $\tau_s$ is \textit{not} a multiple of $\pi$, say
$n\pi < \tau_s < (n+1)\pi$, the Hopf bifurcation to the internal mode still
occurs before the bifurcation to any of the external modes. The frequency of the internal mode
is $\omega \approx 1$ (instead of  $\omega =1$), and $\Delta_H > \epsilon^2 c_2$
but still $O(\epsilon^2)$. Thus, if we plot $\Delta_H$ as a function of
the delay $\tau_s$, the \textit{minimum} of the curves $\Delta_H(\tau_s)$
occur when $\tau_s = n\pi$, $n$ odd. This is illustrated in Fig.~\ref{f:linstabfig2}
and will be discussed further below.

Now consider the external modes with $\omega \approx m\pi/\tau_s$. For negative
couplings when $\delta_j > 0$ (see Eq.~(\ref{eq:negcoupling})) and $\Delta > 0$ an analysis of the
characteristic equation shows that $m$ \textit{must be even}. Thus, the
external modes have frequencies
$\omega \approx 2\pi/\tau_s,\;\; 4\pi/\tau_s,$ etc. In Fig.~\ref{f:deltavsomega}
we illustrate three cases when $\tau_s = n \pi$, $n=1,\; 5,$ and $53$.
The external modes then have $\omega = m/n$, $m$ even.

Finally, we consider the case when $\tau_s = n\pi$, for $n$ \textit{even}.
As discussed in the previous paragraph, for negative couplings the external
modes with frequency $\omega = m\pi/\tau_s$ require that $m = 2, 4, 6 \ldots$.
Thus, for each external mode there is a value of $\tau_s = n\pi$, $n$ even,
such that $\omega \approx 1$ similar to the internal mode. This is illustrated
in the numerical data of Fig.~\ref{f:linstabfig2}a for $\tau_s = 2\pi$, while
in Fib.~\ref{f:linstabfig2}b we see that the bifurcations to these
two modes occur at nearly the same
value of the coupling. Referencing Fig.~\ref{f:deltavsomega}, the two modes
would have frequencies on opposite sides of $\omega =1$ such that $\Delta_H$ is the same for each.
This is referred to as a ``double Hopf" \cite{PiErGaKo00,GoGuKh97} and may
lead to more complicated bifurcation scenarios that we will not pursue in
the present paper. We also note that at the double Hopf bifurcation, the
designation of ``internal mode" versus ``external mode" is ambiguous because
both have $\omega \approx 1$. Indeed, following the numerical data in
Fig.~\ref{f:linstabfig2}a, as $\tau_s$ is varied through $2 \pi$,
we see that $n=1$ internal mode becomes the $m=2$ external mode, while
the $m = 2$ external mode becomes the $n=3$ internal mode.

%*******************************************************************************
%*******************************************************************************
\subsection{Approximations to internal and external modes}

As discussed above, for any delay $\tau_s$ when the coupling is increased,
the first periodic solutions to appear correspond to the internal mode and
will have frequency $\omega \approx 1$. For arbitrary delay we can find an
approximation for the frequency by letting
$\omega = 1 + \epsilon \omega_1 + O ( \epsilon^2 )$ in
Eq. (\ref{eq:omega}) to find that $\omega_1$ satisfies
\begin{equation}
  \tan ( \tau_s ) [ 4 \omega_1^2 - c_2 ] - 2 c_1 \omega_1 = 0,
  \label{eq:omegaint}
\end{equation}
which is a simple quadratic for $\omega_1$. The bifurcation point is then
approximated as
\begin{eqnarray}
  \Delta_H^2 & = & \epsilon^4 [ 16 \omega_1^4 + 4 ( c_1^2 - 2 c_2 ) \omega_1^2
  + c_2^2 ], \nonumber\\
  & = & \epsilon^4 ( 4 \omega_1^2 + a_1^2 ) ( 4 \omega_1^2 + a_2^2 ) .
  \label{eq:DHint}
\end{eqnarray}

\begin{figure}
\begin{center}
\includegraphics[width=4in]{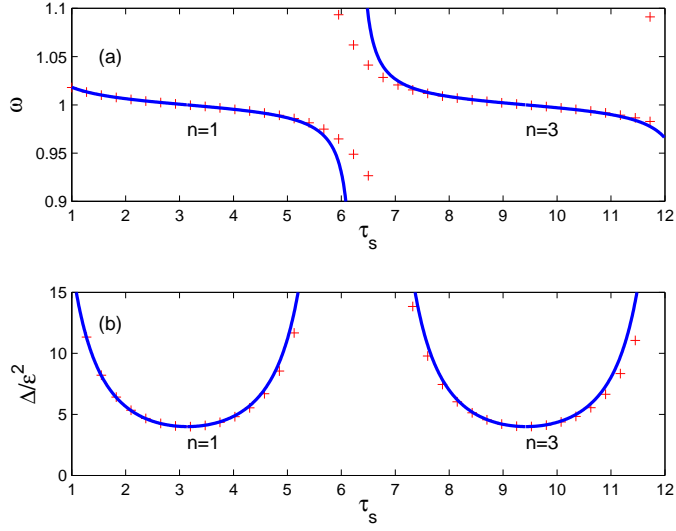}
\end{center}
\caption{Frequency
$\omega$ and coupling $\Delta = \delta_1 \delta_2$ at the Hopf bifurcation
point as a function of the delay $\tau_s = \tau_1 + \tau_2$. The solid curve is
the asymptotic results of Eq. (\ref{eq:omegaint}) and (\ref{eq:DHint}), while the +
are the result of numerically evaluating Eq. (\ref{eq:omega}) and
(\ref{eq:DH}). ($\epsilon = 0.01, a_1 = a_2 = 2, \beta = 1)$}
\label{f:linstabfig1}
\end{figure}

We can obtain simpler results than Eqs. (\ref{eq:omegaint}) and (\ref{eq:DHint})
if we require that $\tau_s \approx n \pi$, $n$ odd; that is, in Fig.~\ref{f:linstabfig1}
we examine locally to the critical values $\omega(\tau_s) = 1$ and
the minimum of the curves $\Delta_H(\tau_s)$.
If $\tau_s = n \pi + \epsilon \tau_{s1}$, then
\begin{equation}
  \omega = 1 - \epsilon^2 \frac{c_2}{2 c_1} \tau_{s} + O ( \epsilon^2 ), \quad
  \Delta_H^2 = \epsilon^4 c_2^2 + O(\epsilon^5).
  \label{eq:omegaDHint}
\end{equation}

\begin{figure}
\begin{center}
\includegraphics[width=4in]{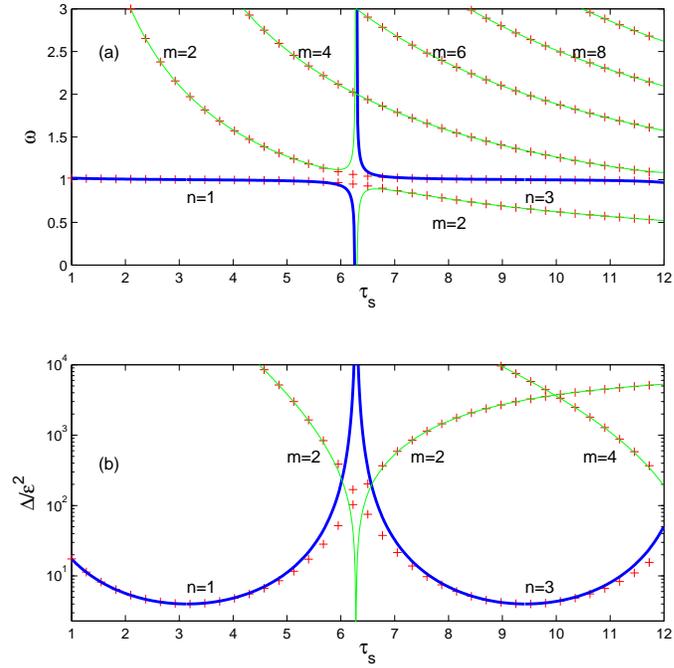}
\end{center}
\caption{Same as Fig.~\ref{f:linstabfig1} but with a wider range in $\omega$ and $\Delta$.
The $+$ indicate the numerical results. The thick curves are the analytical approximations
of the internal modes (same as Fig.~\ref{f:linstabfig1}). The thin curves are the
approximations of the external modes.}
\label{f:linstabfig2}
\end{figure}

For the external modes we let $\omega = \omega_0 + \epsilon \omega_1 + O (
\epsilon^2 )$ with $\omega_0 = m \pi / \tau_s$, $m$ even, and find that
\begin{equation}
  \omega = \omega_0 - \epsilon \frac{\omega_0 c_1}{( 1 - \omega_0^2 ) \tau_s}
  + \ldots
  \label{eq:omegaext}
\end{equation}
and the Hopf bifurcation point is approximated as
\begin{equation}
  \Delta_H^2 = ( 1 - \omega_0^2 )^4 + \epsilon \frac{8 c_1}{\tau_s} ( 1 -
  \omega_0^2 )^2 \omega_0^2 + \ldots
  \label{eq:DHext}
\end{equation}
The external mode results are valid for $\tau_s$ such that $\omega_0 \ne 1$; thus,
$\tau_s$ must be away from the double Hopf point when $\tau_s = m \pi$.

In Figs.~\ref{f:linstabfig1} and \ref{f:linstabfig2} we plot the linear stability given by:
(i) numerically evaluating Eqs.~(\ref{eq:omega}) and (\ref{eq:DH}) (marked with +) and
(ii) the asymptotic approximations from Eqs.~(\ref{eq:omegaint})-(\ref{eq:DHint})
and Eqs.~(\ref{eq:omegaext})-(\ref{eq:DHext}) (solid curves).
Fig.~\ref{f:linstabfig1} focuses on the bifurcation of the internal mode.
For any fixed value of the delay time $\tau_s$, we increase $\Delta$ until the curve
is crossed at $\Delta = \Delta_H$; the frequency is $\omega \approx 1$.
In Fig.~\ref{f:linstabfig2} we pan out so
that both the internal and external modes are visible. From here we see that as
$\Delta$ is increased, the bifurcation to the internal mode always occurs first.
As $\Delta$ is increased further, we cross
secondary curves indicating the bifurcations to the external modes.
More specifically, suppose that $\tau = \pi$ ($n=1$). After the internal mode,
the external modes  with
frequencies $\omega = m/n = m/1$, $m = 2, 4,\ldots$ appear in sequence.
For longer delay times, the
order in which the external modes appear will be different; this can
be seen in Fig.~\ref{f:linstabfig2} for $\tau_s \approx 11$, where the bifurcation
to the external mode with $m=4$ occurs before the mode with $m =2$.
Alternatively, in
Fig.~\ref{f:deltavsomega}b when $\tau_s = 5\pi$, as $\Delta$ is
increased the bifurcation to the periodic solutions occur in the order $\omega \approx 1,\;\;
\approx  0.8,\;\;,\approx  1.2,\;\;, \approx 0.4$, which corresponds to
the internal mode, followed by the external modes $\omega= m/n$, $n=5$, $m=4,6,2,\ldots$.

%*******************************************************************************
%*******************************************************************************
%*******************************************************************************
\section{Hopf Bifurcation of the internal mode}
\label{s:intmodes}

%*******************************************************************************
%*******************************************************************************
\subsection{Two-time scales}

We use the method of multiple scales to analyze the oscillatory solutions that
appear at the Hopf bifurcation points. For the uncoupled lasers, oscillations
decay on an $O ( \epsilon )$ time scale, which suggests that we introduce the
slow time $T = \epsilon t$; time derivatives become $\frac{d}{dt} =
\frac{\partial}{\partial t} + \epsilon \frac{\partial}{\partial T}$. We
analyze the nonlinear problem using perturbation expansions in powers of
$\epsilon^{1 / 2}$, e.g., $x_j ( t ) = \epsilon^{1 / 2} x_{j 1} ( t, T ) +
\epsilon x_{j 2} ( t, T ) + \ldots$; the relevant nonlinear terms and the
slow-time derivative then balance at $O(\epsilon^{3/2})$. Finally, for simplicity
we set $\beta = 1$ indicating identical lasers; however, relaxing this assumption to nearly
identical lasers ($\beta = 1 + O(\epsilon)$) does not qualitatively change the
results.

We now consider the effect of the two-time scale assumption on the delay term.
With the additional slow time, the delay term becomes
\begin{equation}
  y_j ( t - \tau ) \rightarrow y ( t - \tau, T - \epsilon \tau ) .
  \label{eq:delayterm}
\end{equation}
If $\delta_j \ll 1$, then
consideration of the delay term is postponed until higher order and appears as part
of the solvability condition for the slowly varying amplitude. However, if
$\delta_j = O ( 1 )$, then the leading-order problem will contain the delay term
and analytical progress is much more difficult.

If $\tau \gg O ( 1 / \epsilon )$ (large delay), then $\epsilon \tau = O ( 1 )$
and the delay term must be retained in the slow argument. However,
if $\tau = O ( 1 )$, then the slow argument can be expanded as \cite{PiErGaKo00}
\begin{equation}
  y_j ( t - \tau ) = y_j ( t - \tau, T ) - \epsilon \tau
  \frac{\partial}{\partial T} y_j ( t - \tau, T ) + \ldots
  \label{eq:dt-expansion}
\end{equation}
Now the delay of the slow argument is postponed to higher order. We shall see
that this leads to simpler slow-evolution equations for $\delta \ll 1$ (see
Eq.~(\ref{eq:Aonetau})), and is
necessary to make any progress at all for $\delta = O ( 1 )$
(see Sec.~\ref{s:extmodes}).

It should be noted that care must be taken when using a series expansion of
a delay term in a differential equation. The Taylor series may itself be justified
but doing so can change the stability of the differential equation. A simple
example is given in \cite{Dr77}, while \cite{DrSaSl73,ElNo73} provide more
theoretical discussions concerning restrictions on the size of the delay.
In our presentation we will check the validity of our approximations by
comparing our analytical and numerical results.

%*******************************************************************************
%*******************************************************************************

\subsection{Bifurcation equation}

From the linear-stability analysis, we know that the first bifurcation will
be to the internal mode with $\omega \approx 1$ when
$\Delta = \delta_1 \delta_2 = O ( \epsilon^2 )$ (we assume that $\tau_s \ne n\pi$,
$n$ even, and thus do not consider the case of the double-Hopf bifurcation). We
present the case that both coupling
constants are of the same order $\delta_j = \epsilon d_j$. However, as long as
$\Delta = O(\epsilon^2)$ the bifurcation results are qualitatively the same.

Proceeding with the multiple-scale analysis, we find that
at the leading order, $O(\epsilon^{1/2})$, we obtain the solutions
\begin{equation}
	y_{j1}(t,T)=B_{j}(T)e^{it}+c.c.,\quad x_{j1}(t,T)=iB_{j}(T)e^{it}+c.c.,
	\label{e:leadingordery}
\end{equation}
which exhibit oscillations with radial frequency 1 on the $t$ time scale.
To find the slow evolution of $B_{j}(T)$, we must continue the analysis
to $O(\epsilon^{3/2})$. Then, to prevent the appearance of unbounded
secular terms, we determine ``solvability conditions'' for
the $B_{j}(T)$. Due to the scalings of $\delta_j$ and $y_j$, the delay terms
$\delta_j y_j(t-\tau_j,T-\tau_{\epsilon,j})$ are $O(\epsilon^{3/2})$ and
contribute to the solvability condition. The final result is
\begin{equation}
  \frac{\partial B_j}{\partial T} = - \frac{a_j}2 B_j(T) - \frac{i}{6} |B_j(T) |^2 B_j(T) + \frac{i}2
  d_k B_k ( T - \tau_{\epsilon,k} ) e^{- i \tau_k},\label{eq:Alongtau}
\end{equation}
$j, k = 1, 2$, $j \ne k$, and where $\tau_{\epsilon,k} = \epsilon \tau_k$.
The effect of the delay in the slow time appears explicitly in the delay terms
$B_k ( T - \tau_{\epsilon_k} )$. The delay in the fast time has resulted in the
exponential terms $e^{- i \tau_k}$. Eq. (\ref{eq:Alongtau}) is valid for arbitrary
values of the delay (we have not simplified the delay term $B_k(T-\tau_{\epsilon,k})$).
However, according to Eq.~(\ref{eq:dt-expansion}), if $\tau = O ( 1 )$, then
$\tau_{\epsilon,k} =O(\epsilon )$ and delay in the slow time is postponed until higher
order. This results in a simpler solvability condition where the delay in the
slow time does not appear.
\begin{equation}
  \frac{\partial B_j}{\partial T} = - \frac{a_j}2 B_j - \frac{i}{6} |B_j |^2 B_j + \frac{i}2
  d_k B_k e^{- i \tau_k}, \quad j, k = 1, 2, \hspace{0.75em}
  \hspace{0.75em} j \ne k. \label{eq:Aonetau}
\end{equation}

Periodic solutions to the original laser equations, Eqs. (\ref{eq:xymodel}),
correspond to $T$-independent solutions to the solvability conditions. The
conditions are the same for both Eqs. (\ref{eq:Alongtau}) and
(\ref{eq:Aonetau}) because the delay terms in Eq. (\ref{eq:Alongtau}) become
constants. The full $T$-dependent solvability conditions, including delay, are
required only to analyze the stability of the periodic solutions.

To determine the amplitude and phase of the periodic solutions, we let
$B_j ( T) = R_j(T) e^{i \theta_j(T)}$, define the phase difference
$\psi = \theta_2 - \theta_1$ and set the time derivatives to zero. We obtain
\begin{eqnarray}
  0 & = & - a_1 R_1 - d_2 R_2 \sin ( \psi - \tau_2 ), \nonumber \\
  0 & = & - a_2 R_2 + d_1 R_1 \sin ( \psi + \tau_1 ), \nonumber \\
  0 & = & -\frac{1}3 ( R_2^2 - R_1^2 ) + d_1 \frac{R_1}{R_2} \cos (
  \psi + \tau_1 ) - d_2 \frac{R_2}{R_1} \cos ( \psi - \tau_2 ).
\end{eqnarray}
We find that
\begin{equation}
 \tan(\psi) = \frac {a_2 d_2 S_2 \sin( \tau_2 ) -a_1 d_1 S_1 \sin( \tau_1 )  }
              {a_2 d_2 S_2 \cos( \tau_2 ) +a_1 d_1 S_1 \cos( \tau_1) },
	 \label{e:S1S2phase}
\end{equation}
where $S_1 = R_1^2$ and $S_2 = R_2^2$ and the amplitudes are found from
the implicit equations
\begin{eqnarray}
  a_2^2 d_2^2 S_2^2 + a_1^2 d_1^2 S_1^2 + 2 a_1 a_2 d_1 d_2 \cos(\tau_s) S_1 S_2
     - d_1^2 d_2^2 \sin^2(\tau_s) S_1 S_2 &=& 0 \nonumber\\
  d_1 d_2 \sin(\tau_s) S_2 (S_1-S_2) + 3 d_1 d_2 \cos(\tau_s)
   (a_2-a_1)S_1 S_2 && \nonumber\\ + 3(a_1d_1^2 S_1^2 - a_2d_2^2S_2^2)&=&0.
  \label{e:S1S2}
\end{eqnarray}
Of particular note is that just as in the linear-stability analysis, the amplitudes
of the periodic solutions depend on the sum of the delays $\tau_s=\tau_1 + \tau_2$.
This corresponds to the effective round-trip time of information from laser-j returning
to laser-j.

In App.~\ref{s:S1S2} we present explicit solutions of Eq.~\ref{e:S1S2} that specify
$S_j = R_j^2$ as a function of the parameters and the delay $\tau_s$. The expressions
are rather complicated but are easily evaluated numerically. To obtain simpler expressions,
we consider $\tau_s = (n+\xi)\pi$, where $\xi \ll 1$ and $n$ an odd integer. That is, we
tune the delay time $\tau_s$ to be near one of the minimums of the neutral stability
curves in Fig.~\ref{f:linstabfig1}. The bifurcation equation is
\begin{eqnarray}
   R_2^2 &=& 3 (a_1+a_2) \frac{\sqrt{D_1}}{D_2} + \xi Z_{21} + O(\xi^2),
   \label{e:bifequ_int}
\end{eqnarray}
where
\begin{equation}
   D_1 = \frac{d_1d_2}{a_1a_2}-1,\quad
   D_2 = \frac{a_2d_2}{a_1d_1}-1,
   \label{e:D1D2}
\end{equation}
and $Z_{21}$ is specified in App.~\ref{s:S1S2}.

\begin{figure}
\begin{center}
\includegraphics[width=4in]{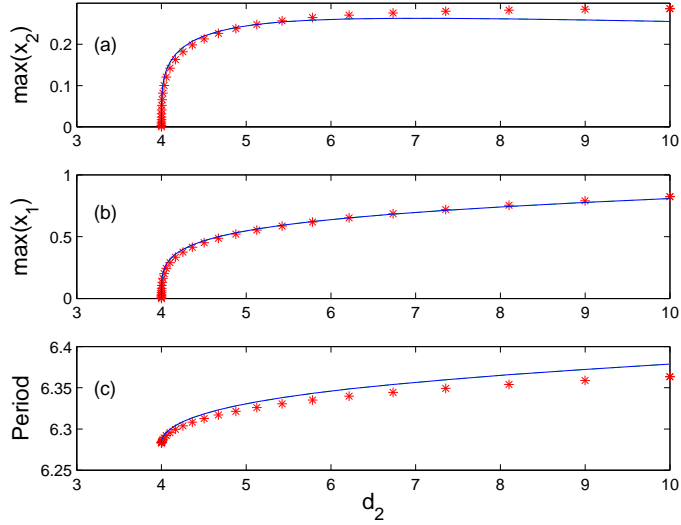}
\end{center}
\caption{Internal mode, $n=5$ with $\tau_1=\tau_2=n\pi/2$, and $\epsilon=0.01$.
   Numerical data points indicated by $*$. The solid curve is the bifurcation
   equation determined by Eqs.~(\ref{e:bifequ_nearmin})-(\ref{e:omega_nearmin})
   for $\xi = 0$, or Eqs.~(\ref{e:R2bifequ})-(\ref{e:R1bifequ}). Because
   $\tau_s=\tau_1+\tau_2 = n\pi$, $\xi=0$ so the bifurcation equations
    are equivalent
   ($d_1 = 1, a_1 = a_2 = 2, b=1, \beta =1$).}
   \label{f:bd_int_n5}
\end{figure}

\begin{figure}
\begin{center}
\includegraphics[width=4in]{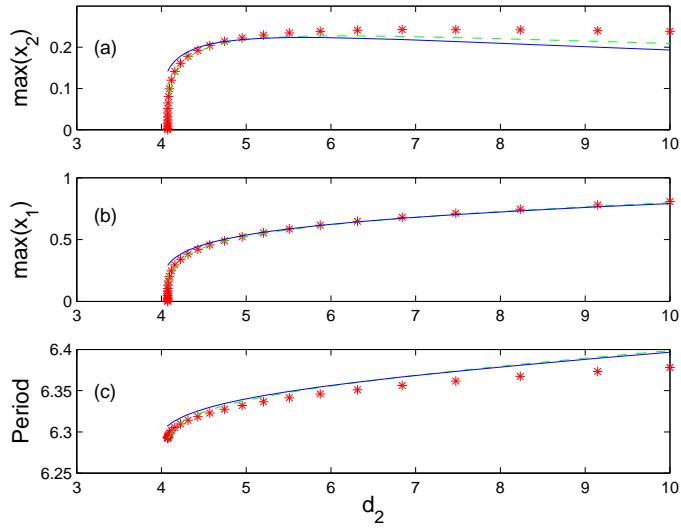}
\end{center}
\caption{Same as Fig.~\ref{f:bd_int_n5} ($n=5$) except that
   $\tau_1=\tau_2=(n+0.1)\pi/2$. Thus, $\tau_s = (n+0.1)\pi$,
   $\xi=0.1\pi$  and
   the delay is tuned to the right of the minimum of the neutral-stability
   curve. The solid curves are the approximate bifurcation
   equations based on $\xi \ll 1$ given by
   Eqs.~(\ref{e:bifequ_nearmin})-(\ref{e:omega_nearmin}). The
   dashed curves are the more general bifurcation equations
   given by Eqs.~(\ref{e:R2bifequ})-(\ref{e:R1bifequ}).
   All other parameters are the same as in Fig.~\ref{f:bd_int_n5}.}
   \label{f:bd_int_n5-2}
\end{figure}

\begin{figure}
\begin{center}
\includegraphics[width=4in]{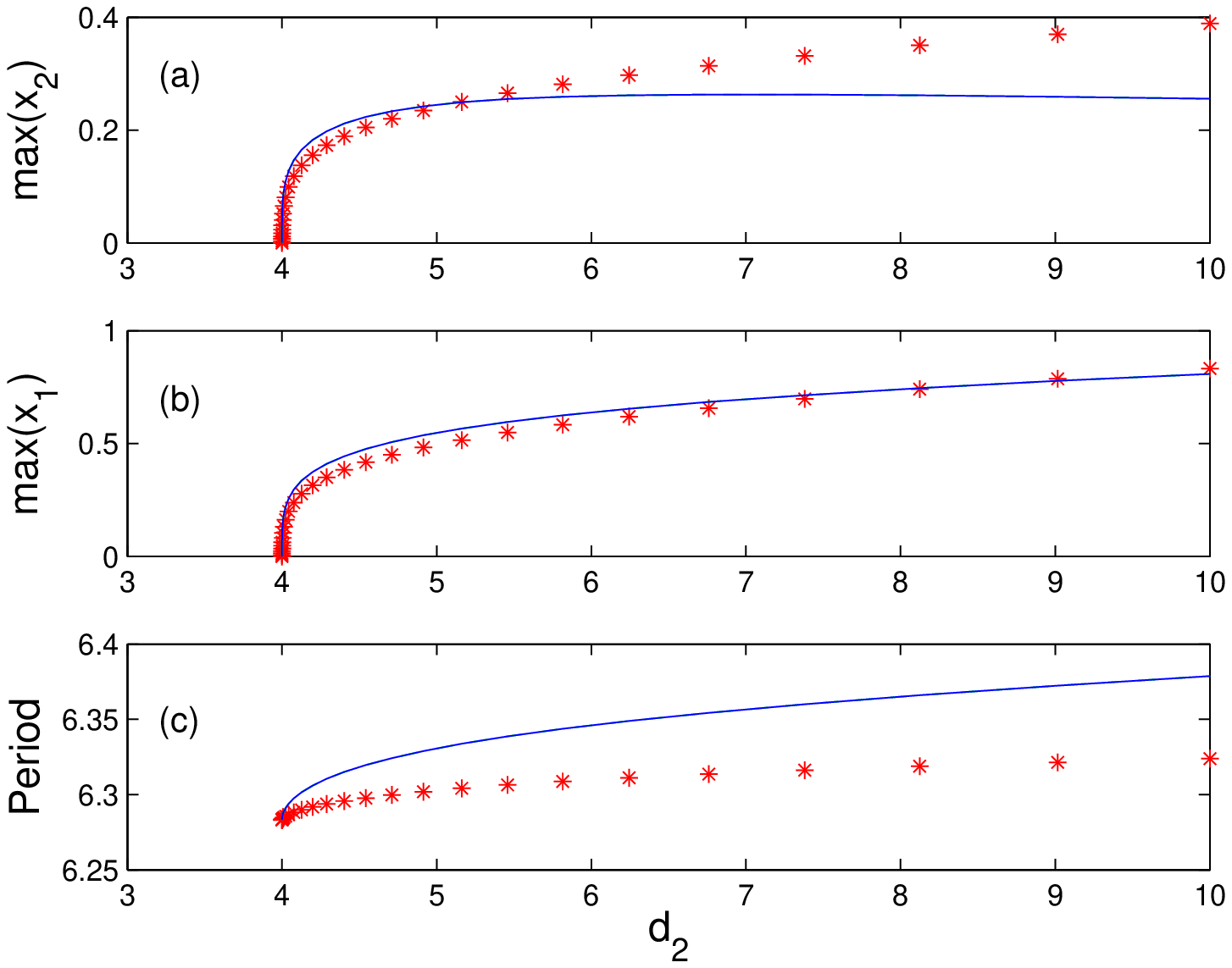}
\end{center}
\caption{Same as Fig.~\ref{f:bd_int_n5} except that $n=53$.}
   \label{f:bd_int_n53}
\end{figure}

We compare the results of numerically computed bifurcation diagrams \cite{ddebiftool} and
our analytical results in Figs.~\ref{f:bd_int_n5}-\ref{f:bd_int_n53}. In Figs.~\ref{f:bd_int_n5}
and \ref{f:bd_int_n53}, $\tau_s = 5\pi$ and $53\pi$, respectively, and hence $\xi = 0$ in
Eq.~(\ref{e:bifequ_int}). The analytical bifurcation equation describes the numerical
results quite well even for $d_2$ far away from the bifurcation point. An expansion
of Eq.~(\ref{e:bifequ_int}) near the bifurcation point shows that
\begin{equation}
   R_2 \sim (\Delta - \Delta_H)^{1/4}, \quad \Delta = \delta_1\delta_2.
   \label{e:internalscale}
\end{equation}


We also note that Eq.~(\ref{e:bifequ_int}) is identical to the bifurcation equation
we derived in \cite{CaTaSc05} for two coupled lasers with opposite sign coupling. The coupled
laser equations in \cite{CaTaSc05} are identical to Eq.~(\ref{e:laserxy}) except that
in \cite{CaTaSc05}: (i) there is no delay such that the coupling is instantaneous and
(ii) we allowed the sign of one of the coupling constants to be positive. Near the
bifurcation point, the small-amplitude periodic oscillations are nearly harmonic
as indicated by Eq.~(\ref{e:leadingordery}). Thus, a positive coupling constant
corresponds to an effective phase shift of half the period or, more generally, a
phase shift of $\tau_s =n\pi$ (see App.~\ref{s:delayandneg}). Thus, we expect
that when $\tau_s$ is tuned to the minimum of the neutral stability curves in
Fig.~\ref{f:linstabfig1}, the bifurcation equations of
the two cases should be equivalent.

In Fig.~\ref{f:bd_int_n5-2} we show the case when $\xi \ne 0$. Here we have tuned
the delay $\tau_s$ to be to the right of the minimum with $\tau_s = (5+0.1)\pi$.
The dashed curve is the result of numerically evaluating the full bifurcation equations,
Eqs.~(\ref{e:S1S2}). The solid curve is the approximation Eq.~(\ref{e:bifequ_int})
and it shows very good agreement with the numerical results.
In general, we maintain good agreement as we tune $\tau_s$ away from the minimum up to
$\tau_s = (n \pm 0.15)\pi$, $n$ odd. The full bifurcation equation, Eq.~(\ref{e:S1S2}),
where we have not made any assumption on the delay,
maintains good agreement out to at least $\tau_s = (n\pm 0.4)\pi$; beyond that however,
it loses fidelity as we approach the double-Hopf bifurcations at $\tau_s = (n\pm 1)\pi$
($\tau_s = n\pi$, $n$ even).

Finally, we note that the first correction to the bifurcation equation Eq.~(\ref{e:bifequ_int})
is linear in $\xi$. Thus, the bifurcation equation is not symmetric about the
minimums of the neutral-stability curves; that is, the amplitude will be larger
on one side of the minimum than the other for a given perturbation of $\Delta$
above the minimun.

%*******************************************************************************
%*******************************************************************************
\subsection{Small-coupling resonance}

The bifurcation equation Eq.~(\ref{e:bifequ_int}) is singular when its denominator
is zero ($D_2=0$), or
\begin{equation}
  \Delta = \Delta_{S} \equiv \frac{a_1}{a_2} \delta_{1}^2
\end{equation}
($\Delta = \delta_1 \delta_2$).
If the parameters are such that the singular point is before the Hopf point,
$\Delta_{S} < \Delta_{H}$, then the singularity can be ignored because the
vanishing denominator will not occur when periodic solutions exist for
$\Delta > \Delta_{H}$. However, if
$\Delta_{S} > \Delta_{H}$, the bifurcation equation will exhibit the pole-type
singularity. Near the singular point, the bifurcation equation
predicts that the amplitude of the oscillations will become large corresponding
to a resonance. The condition that $\Delta_{S} > \Delta_{H}$ corresponds to
$\delta_1^2 > \epsilon^2 a_2^2$. Thus, in laser-2 the increase in the population
inversion provided by pump coupling is greater than the effective losses.
In \cite{CaTaSc05}, when the lasers were coupled without delay,
we investigated this resonance in detail and the bifurcation equation matched
the result of simulations very well. For the present case with delay coupling,
however, we find that there are some dramatic differences.

\begin{figure}
   \begin{center}
      \includegraphics[width=5in]{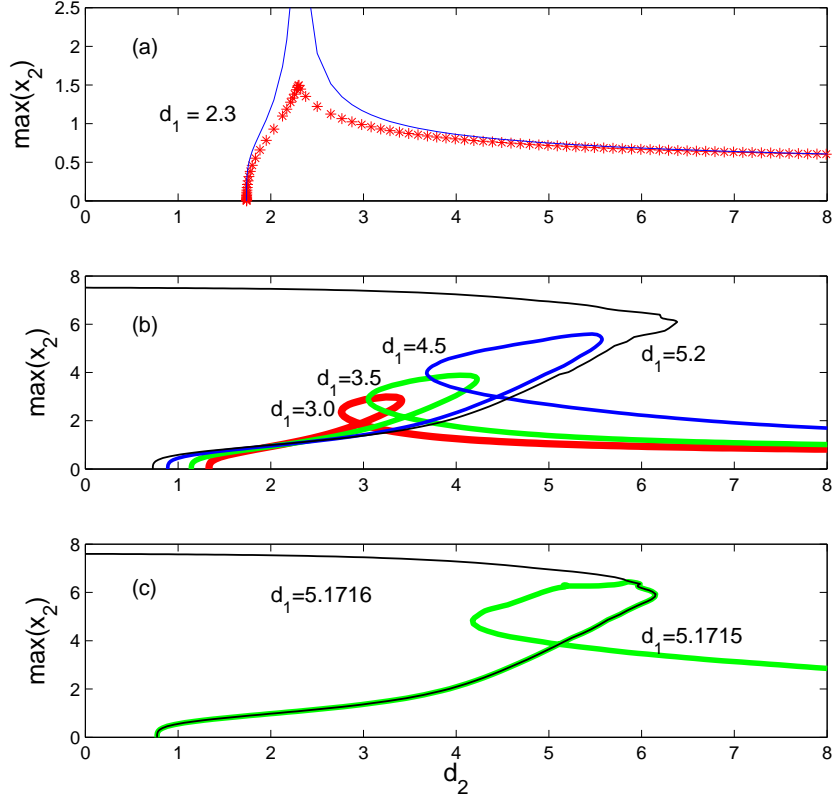}
   \end{center}
   \caption{ Internal mode, $n=1$ with $\tau_s=n\pi$. (a) The solid curve is the predicted
             bifurcation branch from Eqs.~(\ref{e:bifequ_nearmin}) when $d_1$ is tuned
	     such that the singularity in the bifurcation equation occurs after the
	     Hopf bifurcation. The $*$ are the numerical bifurcation data for $d_1=2.3$.
      (b) Numerical bifurcation branches as they deform between $d_1=3.0$ and $d_1=5.2$.
      (c) There is a critical value of $d_1$, where on either side of the critical value
      the bifurcation branch has a loop and proceeds to higher values of $d_1$, or bends
      back without a loop to negative values of $d_1$.
      All other parameters are the same as in Fig.~\ref{f:bd_int_n5}}.
   \label{f:resonance0p010}
  \end{figure}

In Fig.~\ref{f:resonance0p010}a, we compare the analytical and numerical bifurcation
branches. For $d_1=2.3$, the bifurcation equation
predicts a strong resonance peak. The numerical data for $d_1 = 2.3$ ($*$)
exhibits a small resonance peak and there is good agreement between
the analytical and numerical curves both before and after the singularity.
In Fig.~\ref{f:resonance0p010}b we increase $d_1$ from $3.0$ to
$4.5$ to follow the deformation of the bifurcation branch from an isolated
resonance peak to a curve that forms a loop before continuing
to higher values of $d_1$; this folding of the bifurcation branch is new to the
system with delay and was not observed in \cite{CaTaSc05}.
We want to make clear that the bifurcation branch is \textit{not} intersecting
itself. The apparent intersection results from projecting the full bifurcation
curve of $d_1$ versus $(x_1,y_1,x_2,y_2)$, which does not intersect,
onto the $(d_1,x_2)$ plane.

Because the multiple-scale method is a local analysis
and we assumed small-amplitude solutions, it is not unexpected that the
analytical and numerical results match well near the bifurcation point. Similarly,
in the vicinity of the resonance peak, the amplitude of the solutions becomes
$O(1)$ and so it is not surprising that the analytical results fail to match
the numerical results. That said, in our previous work with coupled lasers
without delay \cite{CaTaSc05}, the analytically determined bifurcation equation
matched the numerically computed resonance peak throughout the parameter range,
i.e., even for $O(1)$ amplitudes; that was a bit surprising. We can surmise that
in the present case it is the folding of the bifurcation branch that causes
the mismatch. It may be that if we continued our analysis to higher order and
derived additional nonlinear corrections to the bifurcation equation, we might
better describe the folded branch. This would be a non-trivial and tedious
calculation; but without doing so our supposition remains speculation. In \cite{CaTaSc05}
we used a different asymptotic method better suited to analyzing periodic
solutions with $O(1)$ amplitudes with only partial success.  Its application
to the present problem with delay would be quite complicated and we have not
made the attempt.

In Fig.~\ref{f:resonance0p010}b we increase $d_1$ to 5.0 and see that there
is qualitative change in the bifurcation branch; there is no longer a loop, and
the bifurcation branch continues to negative values of $d_1$. In
Fig.~\ref{f:resonance0p010}b we see that the change in the nature of the bifurcation
curve apppears quite abruptly at a critical value of $d_1$. Typically, one
expects smooth folding and unfolding of bifurcation branches as a parameter is
changed \cite{Wi90}. In the present case, we have not been able to isolate
an interval of $d_1$ over which such an unfolding might occur. To our knowledge,
such a discontinuous unfolding of a branch of solutions in a bifurcation diagram
has not been previously described.

Finally, if $d_1$ is increased further the bifurcation branch folds back and continues
for positive $d_2$; there also appear to be secondary bifurcations but we have not
explored these in any detail. We mention that as the delay $\tau_s$ increases, or as $\epsilon$ increases,
the effect of the singularity is diminished. Conversely, with reduced values of $\epsilon$,
the bifurcation branch may exhibit more complicated turns and folds. It is intuitively
understandable that increasing the damping ($\epsilon$) will diminish any type of resonance
phenomena. The role of the longer delay is unclear.

%*******************************************************************************
%*******************************************************************************
%*******************************************************************************

\section{Hopf Bifurcation of the external modes}
\label{s:extmodes}

In this section, we describe the external modes that emerge via Hopf bifurcations
as the coupling is increased. We fix $\delta_1 = O ( 1 )$ and use $\delta_2$ as the
bifurcation parameter with $\delta_2 = \delta_{20} + \epsilon \delta_{22} + \ldots$.
We again have the slow time $T=\epsilon t$ and the perturbation expansion of $x$ and $y$
in powers of $\epsilon^{1/2}$. The
multiple-scale analysis is more complicated because the $O ( 1 )$
couplings result in a  leading-order problem that contains the delay terms.
Due to the two time scales, the delay terms are of the form
$\delta_j y_j(t-\tau_j,T-\epsilon \tau_j)$. To make analytical progress
we need to remove the slow delay from the leading-order problem. This
requires that we assume $\epsilon \tau_j \ll 1$ such that by using
Eq.~(\ref{eq:dt-expansion}) the slow delay is postponed to higher order.
We then have a restriction on the size of the delay such that
$\tau = o(1/\epsilon)$. Thus, we find that our results fit well with the
external modes corresponding to the case $\tau_s = 5\pi$,
Fig.~\ref{f:deltavsomega}b, but are inaccurate in the case $\tau_s = 53\pi$,
Fig.~\ref{f:deltavsomega}c. Finally, to simplify the presentation, we consider
the case of equal delays $\tau_1 =\tau_2 = \tau$, such that $\tau_s=2\tau$;
analysis of the general case with unequal delays can be carried out in the same way.

%*******************************************************************************
%*******************************************************************************
\subsection{Leading order}

The leading order $\textrm{O} ( \epsilon^{1 / 2} )$ problem is
\begin{equation}
  \frac{\partial}{\partial t} X_1 ( t, T ) = L \cdot X_1 ( t, T ) - D \cdot
  X_1 ( t - \tau, T ), \label{eq:O1DDE}
\end{equation}
where
\begin{equation}
  L = \left( \begin{array}{cccc}
    0 & - 1 & 0 & 0\\
    1 & 0 & 0 & 0\\
    0 & 0 & 0 & - 1\\
    0 & 0 & 1 & 0
  \end{array} \right), \quad D = \left( \begin{array}{cccc}
    0 & 0 & 0 & \delta_{20}\\
    0 & 0 & 0 & 0\\
    0 & \delta_1 & 0 & 0\\
    0 & 0 & 0 & 0
  \end{array} \right),
\end{equation}
\begin{displaymath}
\mbox{and }
 X_1 ( t, T ) = \left(
  \begin{array}{c}
    x_{11} ( t, T )\\
    y_{11} ( t, T )\\
    x_{21} ( t, T )\\
    y_{21} ( t, T )
  \end{array} \right) .
\end{displaymath}
We look for oscillatory solutions of the form $X_1 ( t, T ) = U_1 B ( T ) \exp
( i \omega t )$, where $B ( T )$ is a slowly varying scaler amplitude (to be
determined from a solvability condition at $O ( \epsilon^{3 / 2} )$). $U_1$
is a vector that is determined by substituting our ansatz into Eq.
(\ref{eq:O1DDE}) to obtain
\begin{equation}
  0 = J \cdot U_1 \quad \mbox{where} \quad J = \left( \begin{array}{cccc}
    - i \omega & - 1 & 0 & -\delta_{20} e^{- i \omega \tau}\\
    1 & - i \omega & 0 & 0\\
    0 & -\delta_1 e^{- i \omega \tau} & - i \omega & - 1\\
    0 & 0 & 1 & - i \omega
  \end{array} \right) .
\end{equation}
For a nonzero solution $U_1$, we require $\det J = 0$. This results in the same
condition obtained from the leading-order linear-stability problem;
specifically,
\begin{equation}
  ( 1 - \omega^2 )^2 - \delta_1 \delta_{20} e^{- i 2 \omega \tau} = 0.
\end{equation}
Thus, we have that
\begin{equation}
  \omega = \frac{m \pi}{2 \tau}, \quad m = \mbox{even, positive integer} .
\end{equation}
For later reference it will be useful to note that
\begin{equation}
  e^{- i 2 \omega \tau} = 1, \quad \mbox{and} \quad e^{- i \omega \tau} = \pm
  1 = \nu_m,
\end{equation}
where $\nu_m = + 1$ if $m / 2$ is odd (e.g. $m = 2,6,\ldots )$ and $\nu_m = - 1$ if $m
/ 2$ (e.g. $m=4,8,\ldots$) is even. Finally, we find that
\begin{equation}
  U_1 = \left( \begin{array}{c}
    i \omega\\
    1\\
    i \omega u_1\\
    u_1
  \end{array} \right) \quad \mbox{where} \quad u_1 = \nu_m
  \sqrt{\frac{\delta_1}{\delta_{20}}} .
\end{equation}

%*******************************************************************************
%*******************************************************************************
\subsection{Second order}

At $O ( \epsilon )$ the problem is
\begin{equation}
  \frac{\partial}{\partial t} X_2 ( t, T ) = L \cdot X_2 ( t, T ) - D \cdot
  X_2 ( t - \tau, T ) + F_2,
\end{equation}
\begin{displaymath}
  \mbox{ where} \quad F_2 = \left(
  \begin{array}{c}
    0\\
    x_{11} y_{11}\\
    0\\
    x_{21} y_{21}
  \end{array} \right) .
\end{displaymath}
Because the homogeneous problem is the same as the $O ( \epsilon^{1 / 2} )$
problem we can, without loss of generality, set the homogeneous solution to 0.
The inhomogeneous term $F$ is proportional to $\exp ( i 2 \omega t )$ so that
the solution is
\begin{equation}
  X_2 ( t, T ) = B ( T )^2 U_2 e^{i 2 \omega t} + c.c.
\end{equation}
where $U_2$ is specified in App.~\ref{s:extmode_coeff}.

%*******************************************************************************
%*******************************************************************************
\subsection{Third order }

At $O ( \epsilon^{3 / 2} )$ we find the solvability condition that determines the
slow-evolution equation for $B ( T )$. The $O ( \epsilon^{3 / 2} )$  problem is
\begin{equation}
  \frac{\partial}{\partial t} X_3 ( t, T ) = L \cdot X_3 ( t, T ) - D \cdot
  X_3 ( t - \tau, T ) + F_3, \label{eq:O3DDE}
\end{equation}
where
\begin{equation}
  F_3 = \left( \begin{array}{c}
    - a_1 x_{11} - \delta_{22} y_{21} ( t - \tau,T ) + \delta_{20} \tau
    \frac{\partial}{\partial T} y_{21} ( t - \tau,T ) - \frac{\partial}{\partial
    T} x_{11}\\
    x_{12} y_{11} + x_{11} y_{12} - \frac{\partial}{\partial T} y_{11}\\
    - a_2 x_{21} + \delta_1 \tau \frac{\partial}{\partial T} y_{11} ( t - \tau,T
    ) - \frac{\partial}{\partial T} x_{21}\\
    x_{22} y_{21} + x_{21} y_{22} - \frac{\partial}{\partial T} y_{21}
  \end{array} \right) .
\end{equation}
The vector $F_3$ has terms proportional to $\exp ( i \omega t )$ and $\exp ( i 2
\omega t )$ and the former will lead to solutions of the form $( U_3 + V_3 t )
\exp ( i \omega t )$. The secular terms $V_3 t$ must be eliminated to prevent unbounded
solutions for large $t$, which implies that a solvability condition must be imposed
on $F_3$. The solvability condition is formulated as follows: We
look for a solution to Eq. (\ref{eq:O3DDE}) of the form $X_3 = U \exp ( i
\omega t )$ and at the same time identify the terms $F$ in $F_3$ proportional to $\exp
( i \omega t )$. We then obtain a algebraic system of equations for
the vector $U$ as
\begin{equation}
  0 = J \cdot U + F,
\end{equation}
where
\begin{equation}
  F = \left( \begin{array}{c}
    ( - i \omega a_1 - \delta_{22} u_1 \nu_m ) B - ( i \omega - \delta_{20}
    \tau u_1 \nu_m ) B_T\\
    i \omega ( u_{22} - 1 ) |B|^2 B - B_T\\
    - i \omega a_2 u_1 B - ( i \omega - \delta_1 \tau \nu_m ) B_T\\
    i \omega u_1 ( u_{24} - u_1^2 ) |B|^2 B - u_1 B_T
  \end{array} \right) .
\end{equation}
For $U$ to have a non-zero solution, the Fredholm Alternative requires
that $V^H \cdot F = 0$, where $V$ is the solution to $J^H \cdot V = 0$ (the
superscript $H$ refers to Hermitian). We find that
$V^H = ( u_1, i \omega u_1, 1, i \omega )$, and the resulting condition for the
amplitude $B ( T )$ is
\begin{equation}
  \frac{\partial B}{\partial T} = ( p_l + iq_l ) B + ( p_n + iq_n ) B|B|^2,
  \label{e:solvability_ext}
\end{equation}
where $p_{l,n}$ and $q_{l,n}$ are given in App.~\ref{s:extmode_coeff}.

%*******************************************************************************
%*******************************************************************************
\subsection{Bifurcation equation}

To analyze the solvability condition given by\\
Eq.~(\ref{e:solvability_ext}), we let $B ( T ) = R ( T )e^{i \theta ( T )}$ to obtain
\begin{eqnarray}
  \frac{\partial R}{\partial T} & = & ( p_l + p_n R^2 ) R, \\
  \frac{\partial \theta}{\partial T} & = & q_l + q_n R^2 .
\end{eqnarray}
The equation for $\theta$ determines the frequency correction as a function of
the amplitude. The bifurcation equation is determined by considering
steady-state solutions to the equation for $R$, and we find that
\begin{equation}
  R^2 = - \frac{p_l}{p_n} = - \left( \frac{|\delta_1|}{ r_2 \omega^2 |1-\omega^2|}\right)
  [ \delta_1 \delta_{22} - \frac{2 \omega^2 ( a_1 + a_2 )}{\tau} ] .
  \label{e:bifequ_ext}
\end{equation}
The onset of oscillations occurs when $R = 0$ and determines $\delta_{22}$ at
the Hopf bifurcation point; this result matches exactly that obtained in the
linear-stability analysis.

\begin{figure}
\begin{center}
\includegraphics[width=4in]{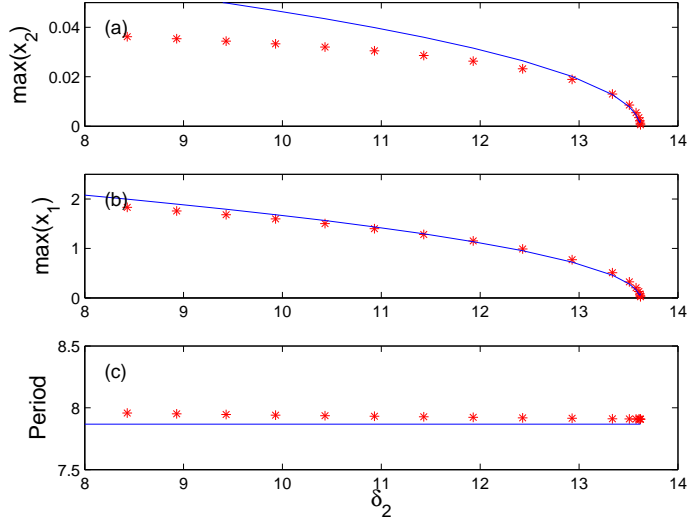}
\end{center}
\caption{External mode, $n=5$, $m=4$ with $\tau_1=\tau_2=n\pi/2$, and $\epsilon=0.01$.
   Numerical data points are indicated by $*$. The solid curve is the asymptotic
   approximation based on Eq.~(\ref{e:bifequ_ext}).
   ($d_1 = 1, a_1 = a_2 = 2, b=1, \beta =1$)}
   \label{f:bd_ext_n5m4}
\end{figure}

\begin{figure}
\begin{center}
\includegraphics[width=4in]{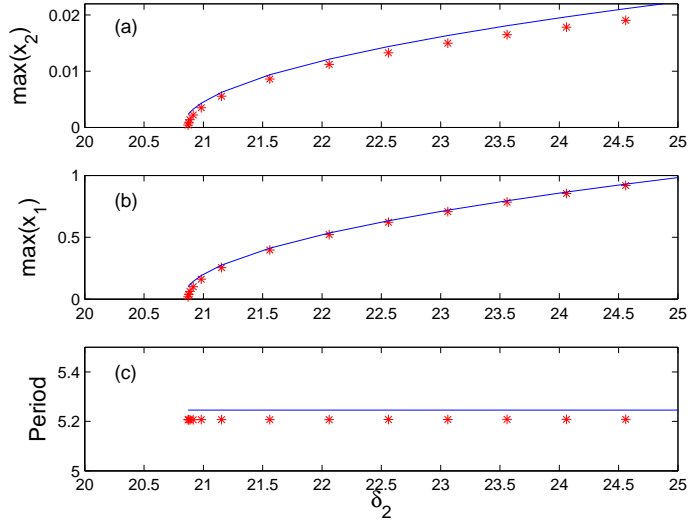}
\end{center}
\caption{Same as Fig.~\ref{f:bd_ext_n5m4} except $m=6$. All other
   parameter values are the same.}
   \label{f:bd_ext_n5m6}
\end{figure}

In Figs.~\ref{f:bd_ext_n5m4} and \ref{f:bd_ext_n5m6}, we compare the bifurcation
equation Eq.~(\ref{e:bifequ_ext}) to the numerically computed result.
In each figure we have $\tau_s = 5\pi = O(1)$ (see Figs.~\ref{f:bd_int_n5} and
\ref{f:bd_int_n5-2} for analysis of the internal mode). In Fig.~\ref{f:bd_ext_n5m4}
the external mode corresponds to $m=4$ and its direction of bifurcation is subcritical.
In Fig.~\ref{f:bd_ext_n5m6} we have $m=6$ and the bifurcation is supercritical. In each
case, there is good agreement between the numerical and analytical result local to the
bifurcation point, where the multiple-scale analysis' validity is strongest.

\begin{figure}
\begin{center}
\includegraphics[height=2in]{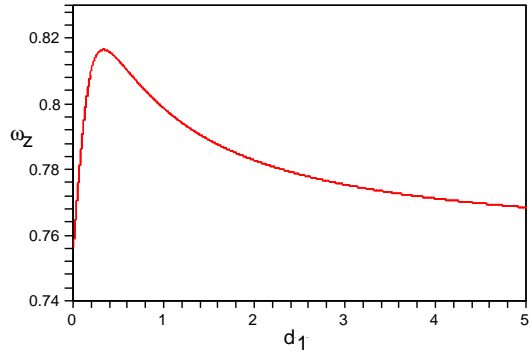}
\end{center}
\caption{Upper bound on the value of $\omega$ as a function of $d_1$
giving a sign change for parameter $r_2$ in Eq.~(\ref{e:r2parameter}).}
\label{f:r2parameter}
\end{figure}

The direction of bifurcation (super- or subcritical) in Eq.~(\ref{e:bifequ_ext}) is
controlled by the sign of the constant $r_2$, which is given in Eq.~(\ref{e:r2parameter}).
Analysis of $r_2$ shows that $r_2>0$ in the interval $\sqrt{2/5} < \omega < \omega_z(d_1)$,
where $\omega_z(d_1)$ is shown in Fig.~\ref{f:r2parameter} (the lower bound is a zero
of the denominator of $r_2$, while the upper bound is the sole real zero of the numerator).
Thus, for external modes with frequencies within this interval, the bifurcation will be
subcritical; this is the case for the external mode in Fig.~\ref{f:bd_ext_n5m4}. For
all other modes $r_2<0$ and the direction of the bifurcation will be supercritical.

\begin{figure}
\begin{center}
\includegraphics[width=4in]{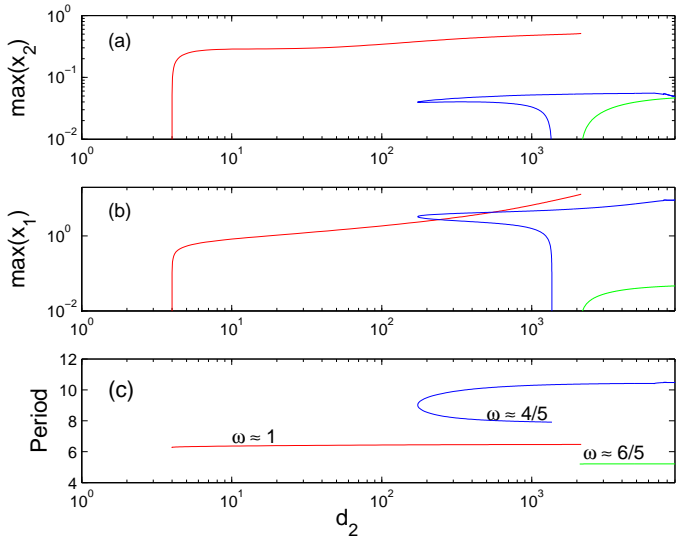}
\end{center}
\caption{Numerically computed bifurcation diagrams, both internal and external modes, for $n=5$
   and $\tau_s = n\pi$. The left-most branch corresponds to the internal mode
   shown in Fig.~\ref{f:bd_int_n5}; periodic solutions are stable until
   $d_2 \approx 59$, when there is a period-doubling bifurcation. The middle and right-most branches
   are the continuation of the external modes shown in Fig.~\ref{f:bd_ext_n5m4}
   ($m=4$) and Fig.~\ref{f:bd_ext_n5m6} ($m=6$), respectively; periodic solutions are unstable
   along these branches
   ($\epsilon=0.01, d_1 = 1, a_1 = a_2 = 2, b=1, \beta =1$).}
   \label{f:bd_intext_n5}
\end{figure}

For reference, in Fig.~\ref{f:bd_intext_n5} we show the bifurcation diagrams of the
internal mode ($\omega\approx 1$) and
the external modes ($\omega = m/n$, $m=4$ and $m=6$) for the case $n=5$; that is, we have
combined the bifurcation diagrams of Figs.~\ref{f:bd_int_n5}, \ref{f:bd_ext_n5m4}
and \ref{f:bd_ext_n5m6}. The subcritical bifurcation for the external mode
$m=4$ folds to provide an interval of hysteresis. The bifurcation branches are projections
onto planes so that the intersections of the curves are not relevant.
The periodic solutions corresponding to the external modes are
unstable as they bifurcate from the unstable branch of steady-state solutions.
The internal mode is stable until $d_2 \approx 59$, when a period-doubling sequence
to chaos begins.

Finally, we note that for the external modes
\begin{equation}
  R \sim [ \Delta - \Delta_H ]^{1 / 2} .
\end{equation}
The amplitude of the external modes varies as the square root
of the distance from the bifurcation point. In contrast, the internal mode
varies as 1/4 the power of the deviation.

%*******************************************************************************
%*******************************************************************************
%*******************************************************************************
\section{Discussion}

We have analyzed the output of two mutually coupled lasers, where the light intensity
deviations from steady state modulated the pump of the other laser. The coupling
strength in one direction is held fixed, while we examine the effect of
increasing the coupling strength in the other direction.
The signal-propagation time through the optical fiber
and the electronic circuit causes a delay leading to a model that is
a system of delay-differential equations.
Linear-stability analysis finds that the steady-state solution becomes unstable
at a Hopf bifurcation; we call the resulting periodic solution the internal mode
because it oscillates at the laser's relaxation-oscillation frequency.
As the coupling is increased,
subsequent instabilities occur with frequencies determined by the round-trip
delay time $\omega_e = m\pi/\tau_s$, $m=2,4,6,\ldots$; we call these periodic
solutions external modes.

Using a multiple-scale analysis, we derive bifurcation equations for both the
internal and external modes. We find that the amplitude of the internal
mode increases with the 1/4-root of the deviation from the bifurcation
point and is supercritical; i.e., $R\sim +(\Delta-\Delta_H)^{1/4}$ .
The amplitude of the external modes increase with the 1/2-root of the deviation and may be super- or subcritical; i.e., $R\sim \pm (\Delta-\Delta_H)^{1/2}$.
Both the initial
instability and the bifurcation results depend on the product of the
coupling constants $\Delta = \delta_1 \delta_2$. For the analysis of the
internal mode, we assumed that both coupling constants were of the same relative
size, $\delta_j = O(\epsilon)$. However, other scalings satisfy the Hopf condition,
e.g., $\delta_1 = O(\epsilon^{1/2})$ and $\delta_2 = O(\epsilon^{3/2})$. We have
found that this does not change the qualitative properties of the bifurcation.

We have focused our analysis on the onset of oscillatory instabilities
and just beyond. We have not considered the stability of the internal and
external modes as the coupling is increased further. However, numerical simulations
indicated period-doubling bifurcations of individual modes as well as
multimode behavior.

With independent control of the individual coupling strengths,
we have observed an atypical resonance phenomena. As in our
previous work on two mutually coupled lasers without delay \cite{CaTaSc05},
there is an interval of the coupling parameter $\delta_2$ over which
the amplitude of the oscillations becomes large; we have referred to
this effect as a ``resonance." The bifurcation equation, Eq.~(\ref{e:bifequ_int}),
can describe the amplitude of the oscillations for $\delta_2$ above
and below resonance. A singularity in the bifurcation equation
indicates the existence of the resonance, but Eq.~(\ref{e:bifequ_int})
does not describe the amplitude within the resonance well.
The parametric form of the singularity indicates that physically, the coupling
term provides an effective negative-damping that cancels with the lasers self-damping
and hence provides a resonance effect.

What is different from the results in \cite{CaTaSc05} is that the bifurcation
branch folds with a change in the coupling parameter $\delta_1$;
in Figs.~\ref{f:resonance0p010} we see that the
resonance peak forms a loop in the $(d_1, x_2)$ plane. We also find that
instead of a smooth unfolding of the loop, there appears to be a critical value of the
parameter $d_1$ where the folded portion of the bifurcation branch
abruptly disappears. To our knowledge, such a discontinuous unfolding of a
branch of solutions in a bifurcation diagram has not been previously described.
We do not understand the mechanism or manner by which the abrupt change
occurs, but this will be the focus of future work.

The application of multiple-scale perturbation techniques to
DDEs is, to our knowledge, a relatively recent development \cite{PiErGaKo00}.
For our analysis of the internal modes, the delay terms appeared only as
part of the solvability condition; thus, the multiple-scale analysis was
relatively straightforward. In contrast, for the external modes that
bifurcate when the coupling constants are $O(1)$, the delay terms are
included in the leading-order problem. However, by looking for periodic
solutions we are able to continue the analysis and formulate a solvability
condition. Finally, as mentioned in the text, multiple-scale expansion of the
delay term in Eq.~(\ref{eq:dt-expansion}) amounts to a Taylor series
expansion that, while it may be justified in and of itself, can lead to
erroneous results for the DDE \cite{Dr77}. Thus, it is important to compare
our analytical results to those from numerical simulations as an important
check of the work.

As discussed in the introduction, the method of averaging has also been used
to analyze the weakly-nonlinear characteristics of delay problems. Averaging
and the multiple-scale technique will lead to a similar slow-time evolution
equation for the amplitudes. However, the multiple-scale technique accounts
for the delay in the slow time where averaging does not. Because we focused
on the existence of periodic solutions, the slow-time delay is removed and
both averaging and multiple-scales give equivalent results. However, stability
of the periodic solutions, or the investigation of more complicated phenomena
such as quasiperiodicity, would require the slow-delay from the multiple-scale
analysis.

Finally, we finish with a discussion relating the results of our analysis
to experimental results observed in \cite{KiRoArCaSc05} and \cite{Ki05}.
The coupling circuit in the experimental system
has two important characteristics. First, the signal is inverted, which results in
the negative signs that appear in front of the coupling constants in Eq.~(\ref{e:laserxy}).
Second, the circuit acts as a low-pass filter that suppresses coupling of the relaxation-oscillations. We discuss the effect of each of these below.

The signal inversion of the coupling circuit results in negative coupling constants
in Eq.~(\ref{e:laserxy}).
However, both the linear-stability and the leading-order bifurcation results depend on
the product of the coupling constants $\Delta = \delta_1\delta_2$. Thus, local to
the Hopf bifurcation it does not matter if both $\delta_j$ are negative
or both are positive. More generally, our results depend only on whether $\Delta$ is
positive or negative, not on the signs of the individual coupling constants. This is
effectively a symmetry result, because for small amplitudes the oscillations
are nearly harmonic. This means a sign change is merely a phase shift.
However, we have observed in numerical simulations that when the amplitudes
become larger such that the intensity is pulsating, the symmetry is lost such
that positive coupling results in different system output from negative coupling.

As mentioned above, the optoelectronic coupling circuit in \cite{KiRoArCaSc05} and \cite{Ki05}
acts as a low-pass filter on the coupling signal. We do not account for this in
our model of the system (recently, Illing and Gauthier \cite{IlGa05} have analyzed a DDE
where they explicitly account for the bandlimited response of their feedback system).
However, because our results address the linear and nonlinear dynamics of both
the internal and external modes, we can make a comparison between theory and
experiment.

\begin{figure}
\begin{center}
\includegraphics[width=4in]{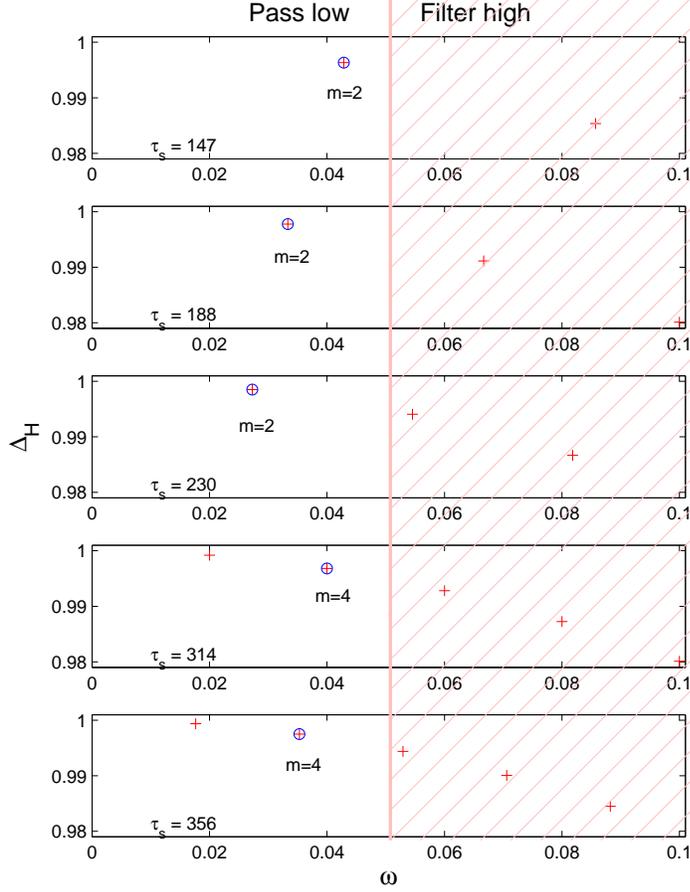}
\end{center}
\caption{For each value of delay, linear stability predicts a set of external modes
with frequencies $\omega_e = m\pi/\tau_s$, $m$ even, indicated by $+$. The frequencies of the
external modes have been plotted on the
linear-stability curve $\Delta_H(\omega)$ as in Fig.~\ref{f:deltavsomega}. External
modes with higher frequency bifurcate first because $\Delta_H(\omega)$ is
decreasing. For each value of the delay, the ($\circ$) indicates the specific
external mode observed in experiment.
All of the observed modes are less than a cutoff frequency due to the low-pass
filter characteristic of the coupling circuit.}
   \label{f:pervsdel}
\end{figure}

The low-pass filter characteristic of the experimental coupling circuit
attenuates the high-frequency relaxation-oscillations, corresponding to the
internal mode. The result is that the
experimental system oscillates at one of the low-frequency external modes.
This is illustrated in Fig.~\ref{f:pervsdel}, where we indicate the
experimentally observed external mode with a ($\circ$) for different values
of the delay. The ($+$) indicate the theoretical value of $\Delta_H(\omega)$
for a bifurcation to an external mode. Linear stability predicts
that as the frequency increases the coupling strength, $\Delta_H$, required for
a Hopf bifurcation decreases (see also Fig.~\ref{f:deltavsomega} for
small $\omega$). However, there is a filter-cutoff frequency $\omega_c$ such
that the relaxation oscillations and higher-frequency external modes are attenuated
and only the external modes with frequencies $\omega_e < \omega_c$ are observed.

For example, in \cite{KiRoArCaSc05} we observed oscillations with period equal
to the round-trip delay time; specifically, $\omega = 2\pi/T = 2\pi/\tau_s$ corresponding
to the external mode with $m=2$. This indicates that the filter-cutoff frequency, $\omega_c$,
is such that $2\pi/\tau_s <\omega_c < 4\pi/\tau_s$. If $\omega_c$ was greater than
$4\pi/\tau_s$ then we would have expected to observe the external mode with $m=4$
because the value of the coupling at the Hopf-bifurcation point decreases as
$\omega_e$ increases, i.e., it will bifurcate at a lower value of the coupling.

More recent experiments \cite{Ki05} confirm that
with longer delays, external modes with $m>2$ are exhibited; this is shown in the lower
two plots in Fig.~\ref{f:pervsdel}. That is, as the
delay increases, the external modes with larger $m$ will be below the cutoff
frequency,  $\omega_e = m\pi/\tau_s < \omega_c$. And it is always the mode with
largest frequency, but still below cutoff, that is exhibited because
$\Delta_H(\omega)$ is least for that frequency. Taking all five plots together,
the experimental data suggest that the low-pass filter cutoff frequency is
$\omega_c \approx 0.05$ because all of the observed modes have frequency less than
$\omega_c$. Finally,  we add that it was observed in \cite{KiRoArCaSc05}
that the amplitude of the oscillations followed a square-root power law as a function
of the coupling. This is exactly as predicted by the bifurcation equation
Eq.~(\ref{e:bifequ_ext}) for the external modes.

Our comparison between the theory and the experiment does have limitations.
We do not know the detailed filter characteristics of the optoelectronic
coupling circuit. It is known that the filter
profile is most certainly not a step function but is instead frequency dependent;
it depends on the properties of the optical cable, the electronic
coupling circuit and the frequency response of the laser to the modulated pump.
In addition, for large values of the delay the frequency difference between the
external modes becomes very small and many external modes are excited at nearly
the same value of the coupling. For these cases the linear stability theory
may be insufficient to identify the mode that is observed in the experiments.

\section*{Acknowledgments}
I.B.S acknowledges the support of the Office of Naval Research.

%*******************************************************************************
%*******************************************************************************
%*******************************************************************************
\appendix

\section{Amplitudes of periodic solutions}
\label{s:S1S2}

\begin{equation}
   R_2^2 = \frac{N}{D},
   \label{e:R2bifequ}
\end{equation}
where
\begin{eqnarray}
N &=& 3Q \left[a_1d_1^2d_2 \sin^2 \tau_s-a_1^2 d_1(a_1+a_2) \cos \tau_s  \right]\\
&& -3\,a_1d_1^3 d_2 ^2 \sin^3\tau_s
  + 3 a_1^2 d_1^2d_2 (3 a_2 +a_1  ) \sin \tau_s  \cos\tau_s \nonumber\\
&&  + 6a_1^3 a_2 d_1(a_1+ a_2)\sin \tau_s,\nonumber\\
  D &=& Q \sin\tau_s\left[ \,a_1^2d_1 -\,d_1 d_2^2\sin^2 \tau_s +
     2\,a_1 a_2 d_2 \cos \tau_s \right]\\
&&  +d_1^2 d_2^3 \sin^4 \tau_s
  -4\,a_1 a_2 d_1 d_2^2 \cos \tau_s \sin^2 \tau_s
  -4\,a_1^2 a_2^2 d_2 \sin^2\tau_s, \nonumber\\
&&  + 2\,a_1^3 a_2 d_1 \cos\tau_s
  - a_1^2 d_1^2 d_2 \sin^2\tau_s
  +2\,a_1^2 a_2^2 d_2 \nonumber\\
  Q &=& \sqrt{
         d_1^2 d_2^2\sin^2\tau_s
	 -4\,a_1 a_2 d_1 d_2 \cos \tau_s
        -4\,a_1^2 a_2^2
	},
\end{eqnarray}
and
\begin{equation}
R_1^2 = \frac{d_2}{2 a_1^2 d_1}
      \left(d_1d_2\sin^2\tau_s -2\,a_2a_1\cos \tau_s  -Q\sin\tau_s
      \right) R_2^2.
      \label{e:R1bifequ}
\end{equation}

When the delay is tuned to be near one of the minimums of the neutral
stability curves, i.e., $\tau_s = m\pi + \xi$, $\xi \ll 1$, we have
\begin{eqnarray}
   R_2^2 &=& Z_{20} + \xi Z_{21} + O(\xi^2), \nonumber\\
   Z_{20}&=& 3 (a_1+a_2) \frac{\sqrt{D_1}}{D_2},\nonumber\\
   Z_{21}&=& \frac{3(D_1+1)}{2 d_1^2 D_2^2}
             \left[ D_1(a_2^3+3a_2^2a_1)+(a_2-a_1)(d_1^2-a_2^2)\right],\nonumber\\
   D_1 &=& \frac{d_1d_2}{a_1a_2}-1,\nonumber\\
   D_2 &=& \frac{a_2d_2}{a_1d_1}-1.
   \label{e:bifequ_nearmin}
\end{eqnarray}
\begin{equation}
   R_1^2 =\left(\frac{a_2}{d_1}\right)^2 (D_1+1)\left[Z_{20} + \xi (Z_{20}\sqrt{D_1}+Z_{21})\right]+O(\xi^2),
\end{equation}
\begin{equation}
   \omega = 1 - \epsilon\left(\frac{1}{6}Z_{20} + \frac{1}{2}a_2\sqrt{D_1}\right)
    - \xi\epsilon \left[\frac{1}{6}Z_{21}+\frac{1}{4}a_2(D_1+1)\right]+O(\xi^2).
    \label{e:omega_nearmin}
\end{equation}

%*******************************************************************************
%*******************************************************************************
%*******************************************************************************
\section{Delay and opposite-sign coupling}
\label{s:delayandneg}

Consider a negative coupling constant such that
\begin{equation}
   \frac{dx_1}{dt} \sim +\delta_2 y_2(t).
\end{equation}
If $y_2$ is harmonic with frequency $\omega =1$ and given by $y_2 = A \cos(t)$,
then
\begin{eqnarray}
   \frac{dx_1}{dt} &\sim& +\delta_2 A \cos(t),\nonumber\\
                 & &    - \delta_2 A \cos(t - n\pi),\quad n \mbox{ an odd integer},\nonumber\\
		 & &    - \delta_2 A \cos(t - \tau_s),
\end{eqnarray}
where $\tau_s = n\pi$ is the total ``round-trip" or system-delay because the
other coupling constant, $\delta_1$, is negative.

%*******************************************************************************
%*******************************************************************************
%*******************************************************************************
\section{External modes: coefficients}
\label{s:extmode_coeff}

\subsection{Second order: $O(\epsilon)$}
\begin{equation}
  U_2 = \left( \begin{array}{c}
    i \omega ( 2 u_{22} - 1 )\\
    u_{22}\\
    i \omega ( 2 u_{24} - u_1^2 )\\
    u_{24}
  \end{array} \right)
\end{equation}
and
\begin{equation}
  u_{22} = - \frac{2 \omega^2 [ 1 - 4 \omega^2 - \delta_1 ]}{( 1 - 4 \omega^2
  )^2 - \delta_1 \delta_{20}}, \quad u_{24} = - \frac{2 \omega^2 [ u_1^2 ( 1 -
  4 \omega^2 ) - \delta_1 ]}{( 1 - 4 \omega^2 )^2 - \delta_1 \delta_{20}} .
\end{equation}

\subsection{Third order: $O(\epsilon^{3/2})$}
\begin{equation}
  p_l = r_1 [ \delta_1 \delta_{22} \tau - 2 \omega^2 ( a_1 + a_2 ) ], \quad
  q_l =  r_1 \{ \nu_m u_1 \omega [ ( a_1 + a_2 ) \frac{\tau}{\delta_{20}} + 2
  \delta_{22} ] \}
\end{equation}
\begin{equation}
  p_n =  r_1 r_2 \tau \nu_m \frac{\omega^2}{u_1} \delta_1,\quad
  q_n = r_1 r_2 2 \omega^3,
\end{equation}
\begin{equation}
  r_1 = \frac{1}{2 ( \delta_1 \delta_{20} \tau^2 + 4 \omega^2 )}, \quad
  r_2 =\frac{
         \delta_1 ( \delta_1 + \delta_{20} + 4 \omega^2 ) - ( 1 + u_1^2 ) ( 1 - 2
         \omega^2 ) ( 1 - 4 \omega^2 )}
	{ (1-4\omega^2)^2 - \delta_1 \delta_{20} }
  \label{e:r2parameter}
\end{equation}

%*****************************************************************************
%*****************************************************************************
%*******************************************************************************

%\newpage
%\bibliographystyle{unsrt}
%\bibliography{/home/carr/Tex/twcarr,/home/carr/Tex/library}

\end{document}